\documentclass[aps,reprint]{revtex4-2}
\usepackage[utf8]{inputenc}
\usepackage[T1]{fontenc}
\usepackage[english]{babel}
\usepackage{graphicx}
\usepackage[caption=false]{subfig}
\usepackage[colorlinks=true, citecolor=blue, linkcolor=blue, urlcolor=blue]{hyperref}
\usepackage{amsmath}
\usepackage{amssymb}
\usepackage{natbib}
\usepackage{mathrsfs}
\usepackage{array}
\usepackage{multirow}

\DeclareMathOperator*{\Tr}{Tr}

\newcommand{\LinearChainTwoQB}{\ensuremath{%
  \includegraphics[height=1ex]{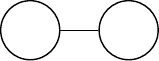}}}

\newcommand{\LinearChainThreeQB}{\ensuremath{%
\includegraphics[height=1ex]{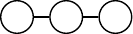}}
}

\newcommand{\VerticalTwoQB}{\ensuremath{\includegraphics[height=1.5ex]{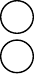}}}

\newcommand{\TriangleOneTwoQB}{\ensuremath{\includegraphics[height=1.5ex]{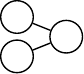}}}

\begin{document}

\title{Error mitigation of entangled states using brainbox quantum autoencoders }


\author{Jos\'ephine Pazem}
\affiliation{Peter Gr\"unberg Institute, Forschungszentrum J\"ulich, J\"ulich 52428, Germany}
\affiliation{Institute for Quantum Information,
RWTH Aachen University, D-52056 Aachen, Germany}

\author{Mohammad H. Ansari}
\affiliation{Peter Gr\"unberg Institute, Forschungszentrum J\"ulich, J\"ulich 52428, Germany}

\begin{abstract}
Current quantum hardware is subject to various sources of noise that limits the access to multi-qubit entangled states. Quantum autoencoder circuits with a single qubit bottleneck have shown capability to correct error in noisy entangled state. By introducing slightly more complex structures in the bottleneck, the so-called brainboxes, the denoising process can take place faster and for stronger noise channels. Choosing the most suitable brainbox for the bottleneck is the result of a trade-off between noise intensity on the hardware, and the training impedance. Finally, by studying R\'enyi entropy flow throughout the networks we demonstrate that the localization of entanglement plays a central role in denoising through learning.
\end{abstract}

\maketitle

\section{Introduction} 

Classical machine learning methods (ML) can identify features in statistics of data and reproduce them \cite{Bishop2006, GoodfellowBengioCourville2016}. Measuring entangled states on quantum systems can sample classical data out of complex probability distributions \cite{Martinis2019}. Recognition of statistical patterns in such data is challenging for classical methods.
Therefore, quantum machine learning techniques (QML) may accelerate or enable the processing of these distributions to recognize such statistical patterns \cite{Schuld2015, Schuld_QMLbook2019, Wittek2017,Marquardt2021}. The quantum speedups, however, can only be characterized in perfect gates, perfect states and  measurements; none of which are so perfect in state-of-the-art devices \cite{Kim2022, Huang2020, Kim2021,Kim2021a}.

 The Noisy Intermediate Scale Quantum (NISQ) processors \cite{Preskill2018} due to their fragility can be reasonably controlled only in the scale of a few tens of qubits \cite{googleAI2023,IBM64V}.  In the absence of fault-tolerant processors, error mitigation requires in-depth characterization of the device and post-processing \cite{Kim2021,Sheldon2021}. Such improvements takes place in the classical simulations of quantum circuits. 

Error mitigation of multi-qubit states is an active research topic and can be approached by increasing coherence time of qubits \cite{Siddiqi2021} or making their net interaction free from unwanted crosstalks \cite{Xu2021, Ku2020}. These attempts will be useful when multi-qubit states are achieved with high fidelity. A crucial task towards this goal is to show the availability of quantum resources such as entanglement on a device \cite{Mermin1990,Mermin1990b,Alsina_Latorre_2016,Huang2020_Mermin}. The power of QML can be leveraged to address the noise impinging on quantum processors. Quantum Neural Networks can contribute to perfecting qubit states on NISQ processors. For this purpose, training tailors the network map to withstand noise and recover the desired quantum features. They are thereby candidates to prove a quantum advantage on near-term devices \cite{Dahlsten2017,Farhi_Neven2018, Cerezo2021_VQAreview,Schuld2021,Beer2020,Cerezo2020_DQNN}.

Autoencoders are a type of neural network that enable the compression of information in smaller layer, the latent space, between input and output layers and are often used to denoise information \cite{Gehring2013,Vincent2008,Vincent2010_StackedAE,Hinton_Salakhutdinov_2006,Sinha2018_birds}. Quantum autoencoders (QAEs) can tackle the problem of producing ideal states using real-device noisy quantum gates \cite{Cao2020,Romero2017,Bravo-Prieto_2021,Du2021,Bondarenko2020}. Noisy devices are unable to prepare ideal entanglement. QAEs, however, can be trained on noisy data in an unsupervised fashion so that they produce ideal states as outputs. To verify this concept, an autoencoder with a single-qubit latent space is trained to reconstruct a perfect Greenberger-Horne-Zeilinger (GHZ) state \cite{Greenberger_Horne_Zeilinger_1989}, i.e. $(|0\rangle^{\otimes m} + |1\rangle^{\otimes m})/\sqrt{2}$ in the presence of random bit and phase flip as well as small unitary noise \cite{Bondarenko2020}.

\begin{figure}[t]
\includegraphics[width=0.47\textwidth]{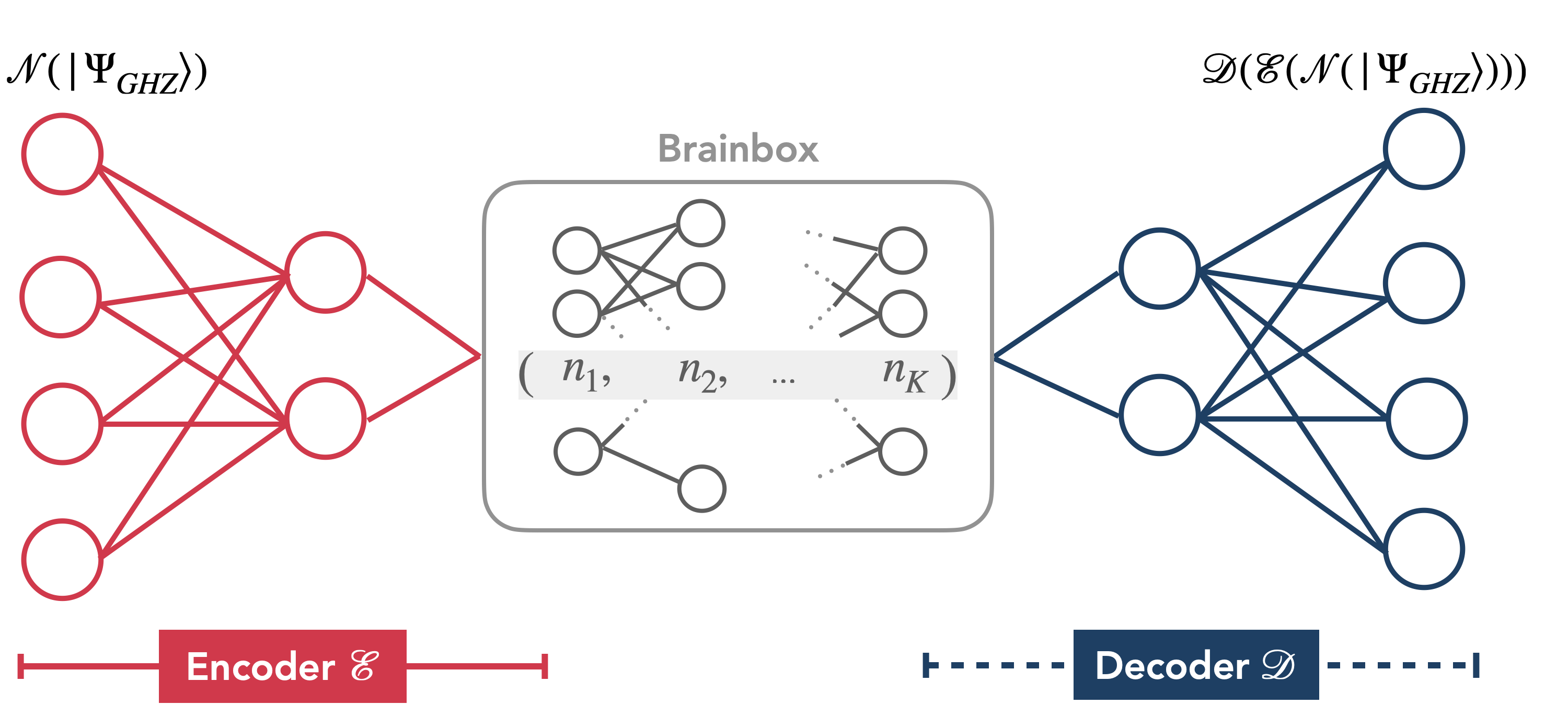}
\caption{Architecture of the brainbox quantum autoencoder with symmetric four-qubit inputs/output layers. The left (red) partition of the network is the encoder, where information of the input is compressed until the brainbox bottleneck by reducing the number of qubits. The right (blue) partition is a decoder that reconstitutes the inputs on the output layer. The brainbox is represented by the set of qubit numbers in a row from left to right, i.e. $(n_1, \cdots, n_K)$. For example, we denote (1,1,1) the $\LinearChainThreeQB$-QAE,(2) the $\VerticalTwoQB$-QAE and (1,2) the $\TriangleOneTwoQB$-QAE.}
\label{fig. network structure}
\end{figure}

In this paper, we use brainbox QAE (BB-QAE) as a generalized QAE in which a small network replaces a single-qubit latent space. These brainbox circuits differ by the number of qubits and their layouts. They can be composed of one or many layers. The morphology of the BB-QAE and brainbox are shown in Figure \ref{fig. network structure}. We aim at denoising GHZ states using different brainboxes and we make a close comparison between them. We show that a strong bit-flip noise beyond the tolerance of single-qubit QAE can be well-tolerated by a rather different brainbox. For example, the noise intensities observed on some qubits of the IBM Eagle chip can be counter-acted. Moreover, we study entropy evolution in the neural network and show that entanglement can be rearranged in the network during the training and this is key to the network's success in tolerating strong noisy flips.

\section{Training of the quantum autoencoder} 

Quantum autoencoder (QAE) network consists of a set of interconnected qubits in layers with a bottleneck in the middle (see Fig.\ref{fig. network structure}). The first (last) layer of the network represents the input (output) register. The edges connecting qubits in adjacent layers represent a quantum map from one layer to the next. There is no connection between qubits of the same layer, meaning that they may be independent on the hardware too. The network's bottleneck is a layer with fewer qubits compared to input and output layers. From the input layer to the bottleneck, the encoder selectively retains information from the input layer to build a good encoding in the bottleneck. Initialized in the computational ground state, the decoder recovers the inputs from the state encoded in the bottleneck. Optimization of the encoder's and decoder's maps relies on the comparison between input and evolved states.

Our QAE is a dissipative quantum neural networks (DQNN) organized in $L$ layers. Each layer $l$ contains $N_l$ qubits, and each qubit in layer $l$ is coupled to all qubits in layer $l+1$. Thus, we univocally denote the network's topology as $(N_1, \cdots, N_L)$. In the middle of the symmetric structure of the QAE, we use a small sub-network instead of single-qubit layer and call it brainbox bottleneck (BB). It can be either mirror symmetric as the QAE, or asymmetric, as depicted in Figure (\ref{fig. network structure}). Varying the morphology of BBs helps to understand how the bottleneck's structure impacts outcome results on the output layer. 

The quantum map on the QAE is constructed starting from the input layer and propagates the state forwards, layer by layer, towards the  output layer. The unitary $U^l_j$ acts on all qubits in layer $l-1$ and $j$-th qubit in layer $l$. It changes the state of the $j$-th qubit in layer $l$. Therefore the quantum map that updates qubits in layer $l$ looks like $\mathcal{U}^l \equiv \prod_{j=1}^{N_l}U^l_j$. For example consider there are $N_5$ qubits in layer $5$ and $N_6$ on layer $6$. The density matrix of layer 6 is initialized in the computational ground state $|0\rangle$ and its transformation depends on the state on layer 5, i.e. $\rho_{(6)}=\rm{Tr}_{(5)} \{\mathcal{U}^{(6)} \left(\rho^{(5)} \otimes |0\rangle\langle0|^{\otimes N_{6}}\right) {\mathcal{U}^{(6)}}^\dag\}$. The trace isolates the state on layer $l$ and dissipation equips the network with forgetfulness, a necessary condition to learning \cite{Bondarenko2023}. Therefore one can easily conclude that the output density matrix can be generated as follows: 
\begin{equation}
\rho^{out} = \prod_{l = 2}^{L} \Tr_{(l-1)} \left\{\mathcal{U}^l \left(\rho^{l-1} \otimes |0\rangle\langle0|^{\otimes N_{l}}\right) {\mathcal{U}^l}^\dag \right\}
\end{equation}
with $\rho^{l-1}$ denoting the density matrix of the layer $l-1$ and all qubits in layer $l$ are in the ground state.

The QAE has been trained with a (1)-BB structure to enable the reconstruction of a noise-free multiqubit entangled states \cite{Bondarenko2020}. The study attempts to prepare ideal GHZ-states $|\Psi_{\rm{in}}\rangle = (|00\cdots 0\rangle + |11\cdots 1\rangle)/ \sqrt{2}$. But noisy hardware is simulated by statistically exposing each qubit to the bit-flip noise channel $\mathcal{N}(\rho_{\rm{in}})$ with flip probability $p$:
\begin{align}
\mathcal{N}(\rho_{\rm{in}}) &= \mathcal{E}_{N_{\rm{in}}}(\cdots(\mathcal{E}_1(\rho_{\rm{in}}, p), p)\cdots)
\label{eq: bit-flip channel}
\end{align}
with $\mathcal{E}_i(\rho_{\rm{in}}, p) = (1-p)\, \rho_{\rm{in}} \, + \, p \, X_i \rho_{\rm{in}} X_i$ being the bit-flip channel for the qubit $i$ and $X_i$ being the flip Pauli operator. For single-qubit bottlenecks, Ref\cite{Bondarenko2020} and Ref\cite{Achache2020} show that the noise tolerance of the (1)-QAE is low ($p<0.3$). We continue the analysis on BB-QAEs with larger brainbox bottlenecks.

The quantum map of the BB-QAE is divided in two parts: the encoder and the decoder. In the left wing of the network, the map $\mathcal{E}(\rho^{in})$ of the encoder is applied on the noisy inputs and hidden layers, and compresses states in the latent space, in the brainbox \cite{Cao2020, Romero2017, Bravo-Prieto_2021, Du2021}. For a brainbox bottleneck with $K$ layers, we note different configurations by $(n_1, \cdots, n_K)$. For example (1,1) is a linear chain of two qubits $\LinearChainTwoQB$.

In the right wing of the network, the decoder map $\mathcal{D}$ reconstructs states in their original dimension, thanks to the information encoded in the last layer of the brainbox. The output state is then $\rho^{\rm{out}}_x = \mathcal{D}(\rho^{\rm{latent}}_x) = \mathcal{D}(\mathcal{E}(\mathcal{N}_x(\rho_{\rm{GHZ}}, p))$, where $\mathcal{N}_x(\rho)$ denotes a discrete noise realization $x$ of the bit-flip channel, that is a combination of flipped/not flipped on all qubits of the input layer.

The aim of the quantum map is to make the output quantum state as similar as possible to the ideal target state. In other words, a successful denoising strategy on a QAE should wash out the statistical noise encoded on the input layer from the output state. This can be measured by evaluating the fidelity of the output state $\rho^{\rm{out}}$ with the ideal state $\rho_{\rm{GHZ}}$:
\begin{align}
F_x(\rho^{\rm{out}}_x, \rho_{\rm{GHZ}}) 
=& \langle \Psi_{\rm{GHZ}} | \, \rho^{\rm{out}}_x | \Psi_{\rm{GHZ}} \rangle \nonumber\\ 
=& \Tr\left\{\rho_{\rm{GHZ}}\, \rho^{\rm{out}}_x\right\}.
\label{eq: fidelity}
\end{align}

At each training step $n$, the average fidelity over all $N_{data}$ states $\{\mathcal{N}_x(\rho_{\rm{GHZ}}, p)\}_{x = 1}^{N_{\rm{data}}}$ defines the objective function for the network:
\begin{equation}
F(n) = \frac{1}{N_{\rm{data}}} \sum_{x = 1}^{N_{\rm{data}}} F_x\left(\rho^{\rm{out}}_x(n), \rho_{\rm{GHZ}}\right).
\label{eq: averaged fidelity}
\end{equation}
The maximization of this function instructs the network how to perform its task. First initialized at random, the interlayer unitaries $\{U^l_j\}$ are updated layerwise and iteratively with the parameter matrix multiplication method \cite{Beer2020, Cerezo2020_DQNN}: 
\begin{equation}
U^l_j(n + \varepsilon) \leftarrow e^{i\varepsilon K^l_j(n)} U^l_j(n),
\end{equation}
where $K^l_j(n)$ is the parameter matrix derived from $F$ \cite{Bondarenko2020}.
This update rule is inspired by the gradient descent algorithms \cite{Ruder2016} and understands gradients as the derivative of $F$ with respect to each unitary. After $N_{\rm{it}}$ updates of the quantum map, the objective function converges to 1 if the training is successful or takes smaller positive values otherwise.

\section{Results}

Denoising a four-qubit GHZ state has been previously performed in \cite{Achache2020,Bondarenko2020} on some symmetric input/output QAE network examples, such as a simple single-qubit bottleneck with additional hidden layers (4,2,1,2,4) and without them (4,1,4), and (4,1,4,1,4) network which is twice (4,1,4).   Each training employs 200 GHZ states exposed to noise of certain bit-flip probability $p$. For each $p$ the performance assessment is evaluated by comparing mean fidelity function before and after applying denoising quantum map. These QAEs can ideally denoise GHZ state up to the tolerance noise $p^*=0.3$, see Fig. (2) in Ref.\cite{Bondarenko2020}.  
We use the same scheme on different network topologies, and our aim is to understand under what topology or connectivity conditions noise tolerance $p^*$ can be improved beyond the weak limit of 0.3. This enable QAEs to denoise entangled states under harsh noise exposure. In addition, it gives better prospects to scale up the inputs.

\subsection{Tolerance threshold} \label{subsec:Elevating noise tolerance}

\begin{figure}[h!]
 \subfloat[]{\includegraphics[scale = 0.5]{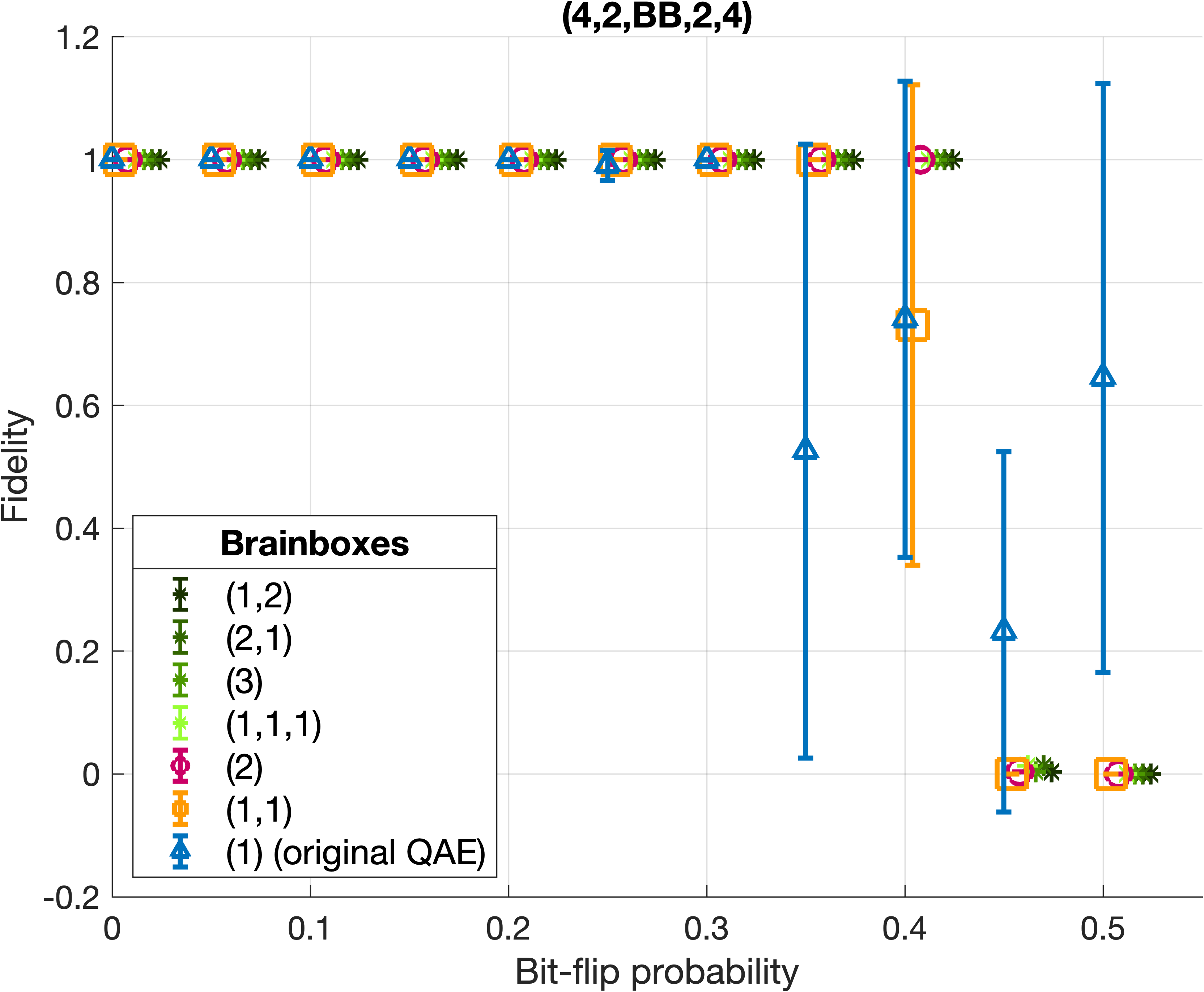}} 
 \centering
 \hfill
  \subfloat[]{\includegraphics[scale = 0.5]{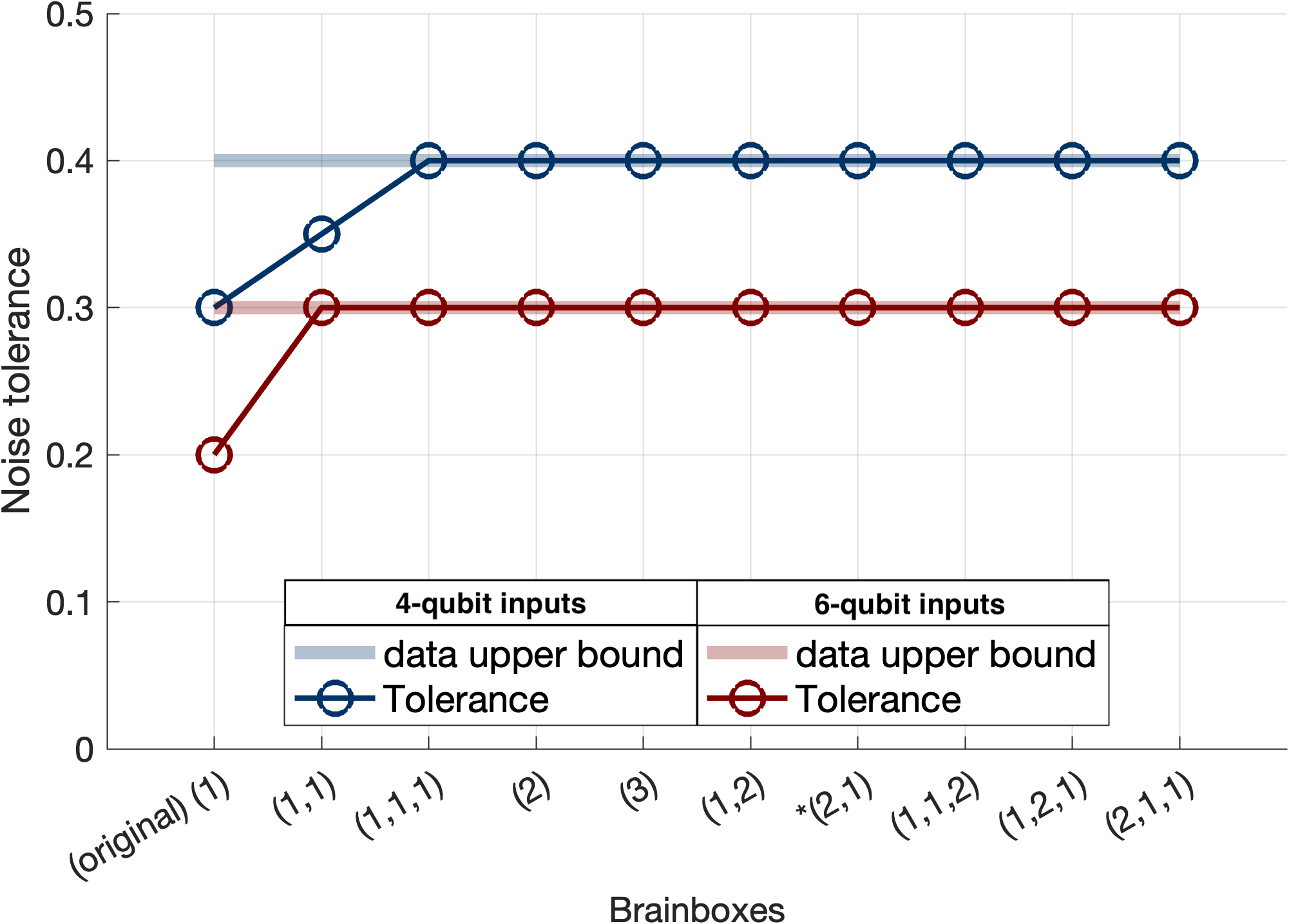} }
    \caption{(a) Testing fidelity: Average output state fidelity over a range of noisy test states with noise probability $p$. The error bars indicates the absolute value of standard deviation in the data about average fidelity. When it is large, it indicates that some noise realizations do not reach high fidelity states after denoising while some do.  
    (b) Tolerance thresholds: The noise probability that returns output states with at least 99\% fidelity with the ideal GHZ state.  Various networks with 4- and 6-qubit input/output layers and different BBs have been tested. Some brainboxes make up to 10\% improvements in the network tolerance threshold.}
    \label{fig:tolerance1}
\end{figure}

Our QAEs contains two symmetric hidden layers each with 2 qubits next to the input and the output, so that the topology is (4,2,BB,2,4) and (6,2,BB,2,6).  We consider the brainbox bottlenecks listed in Fig.\ref{fig:tolerance1}. The BB examples are the symmetric sub-networks (3), (2), (1), (1,1) (1,1,1) and the asymmetric BB sub-networks (1,2) and (2,1).   In order to make a clear comparison between how fast each brainbox makes its way to an optimized quantum map, we start all these networks at the same initial map and we update the map 200 times.  Among these examples, the (1)-QAE represents the original QAE from Ref. \cite{Bondarenko2020}. 

After training the above-mentioned BB-QAEs with training sets of 200 noisy GHZ states and different bit-flip probabilities $p$, testing checks whether the network denoises states it was not trained with. The optimized quantum map is applied to some new noisy GHZ-states. Though the noise realizations differ, the noise channel is the same as during the training. The output states are compared with an ideal GHZ state using the fidelity function \ref{eq: averaged fidelity}.

The result is shown in Fig.\ref{fig:tolerance1}(a). We call tolerance threshold the largest probability the network can recover from, ie the fidelity at the end of the training is close to 1. One can see that the simple brainbox (1) used in Ref. \cite{Bondarenko2020} can tolerate only noise probability up to 0.3, while the (1,1)-QAE can increase the threshold slightly to 0.35. The other brainbox bottlenecks push the tolerance threshold up to 0.4 and perform equally well with respect to fidelity.   

A low tolerance threshold is an issue in two ways. To make the QAE useful on the current NISQ devices, it must operate well at intermediate noise scales, close to the tolerance thresholds. In addition, errors are more likely to occur on larger states. Therefore, noise resilience must be improved. For this aim, we compared the results of (4,2,BB,2,4) and (6,2,BB,2,6) networks. We plot the respective thresholds in Fig.\ref{fig:tolerance1}(b). By increasing the number of input qubits from 4 to 6 while aiming at GHZ-states, the noise tolerance on a simple single-qubit bottleneck shows a large drop off by 0.1 from $p^*=0.3$ to 0.2. This raises concerns about the scalability of denoising: by adding more qubits to the inputs, the noise tolerance shrinks, in other words the QAE becomes more fragile and unable to recover the ideal target state.  

\begin{figure}[h!]
    \centering
\includegraphics[scale = 0.5]{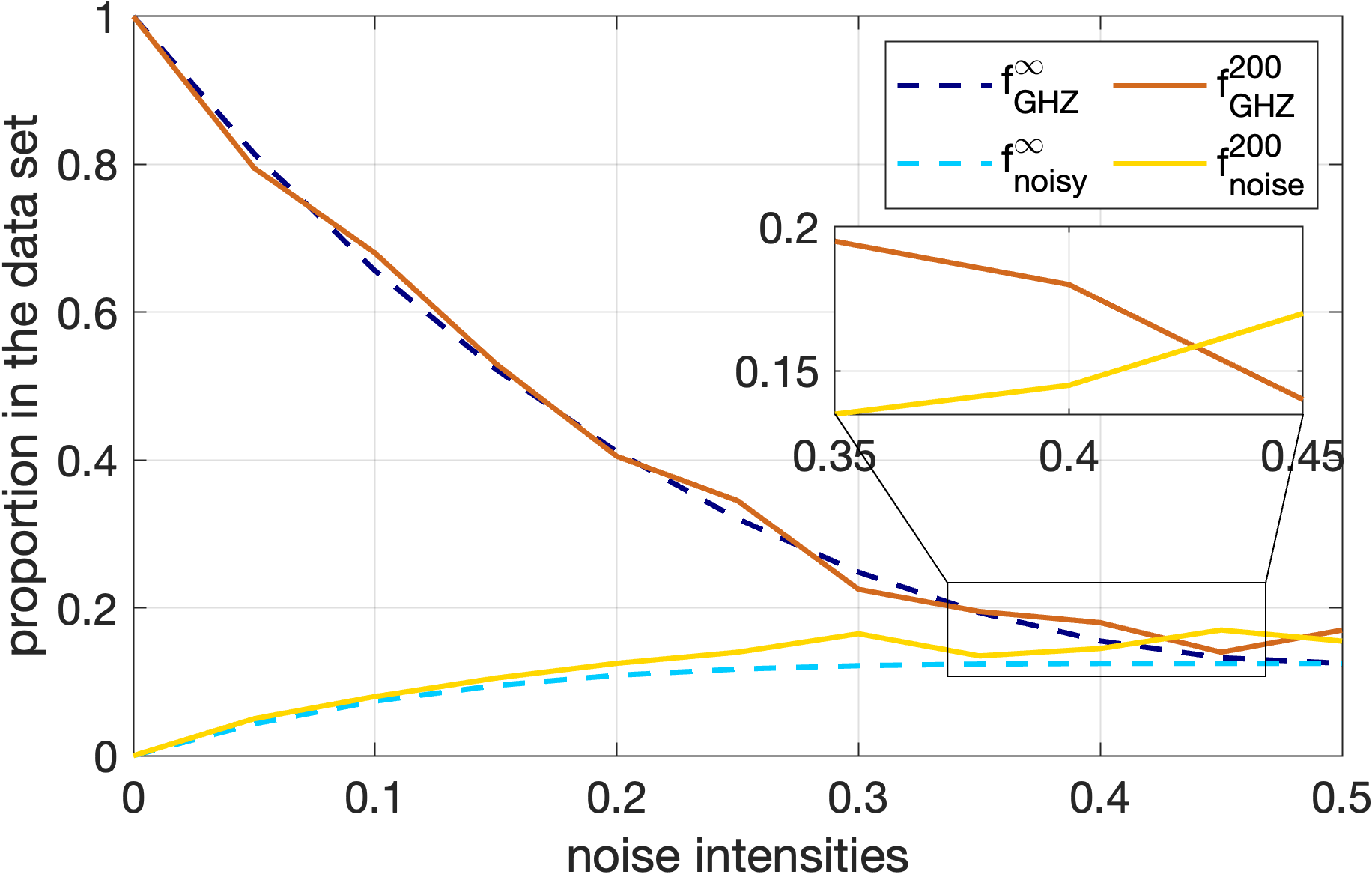}
    \caption{Training data set: The distribution of 4-qubit GHZ and non-GHZ states in infinitely many samples (dashed line) versus finite 200 samples (solid lines). In the infinite sample case the distribution of GHZ for all noise probabilities $p$ dominates, while artefacts in the finite data set prevents the dominance of GHZ for strongly noisy channels.}
    \label{fig:Trainingdataset}
\end{figure}

For a given probability $p$, the number of combinations of flipped/intact qubits in the input states grows exponentially with their size. With a small training data set, this suppresses the tolerance threshold. In the limit of infinite size data set, the distribution of GHZ and non-GHZ states is such that the amount of GHZ states is always larger than that of non-GHZ except at $p=0.5$. With a simple majority rule, the network can identify the GHZ state as the target state. As the size of the training set becomes finite, deviations from the ideal distribution alter the training. Dashed and solid lines in Fig. (\ref{fig:Trainingdataset}) compare distributions for ideal states and the next most probable noisy state for training sets with infinitely many and 200 states respectively. In limited data set, the ideal GHZ state occurs less often in the finite data set than the next most probable noisy state in the vicinity of $p\sim 0.4$. Thus for such data set, the QAE training can only help to boost tolerance threshold up to where GHZ state constitute a majority of the training data. While the training data ultimately imposes an upper bound on the tolerance that can possibly be achieved, the (1)-QAE performs sub-optimally and its tolerance does not depend on the training data. For the multi-qubit BB, he scaling of the generalization error with the size of the training data set is consistent with the results in \cite{Caro2021}.

For this study, a size-200 training set already shows disparities in improvements in tolerance due to the network topology. Fig.\ref{fig:tolerance1}(b) shows that in a network of 4-qubit inputs all brainbox beyond (1) and (1,1) equivalently perform with higher tolerance. Adding two qubits to the input results in the reduction of noise tolerance by 0.1 unit, however in this case still all BBs except the single-qubit bottleneck (1) reveal higher tolerance.  Another important lesson from the study is that qubit configurations in BBs contribute to the tolerance. For example in the case of 4-qubit input, a brainbox with two qubits in separate layers (1,1) yields a sub-optimal tolerance at 0.35, while stacking them in a single layer (2) saturates the data limit at 0.4.

\subsection{Training impedance}

In section \ref{subsec:Elevating noise tolerance}, we found that most of multi-qubit brainboxes we used in the (4,2,BB,2,4) and (6,2,BB,2,6) network maximize the achievable bit-flip noise tolerance $p^*$. However some BB topologies make the training less costly.

\begin{figure}[t]
    \centering
    \includegraphics[scale=0.5]{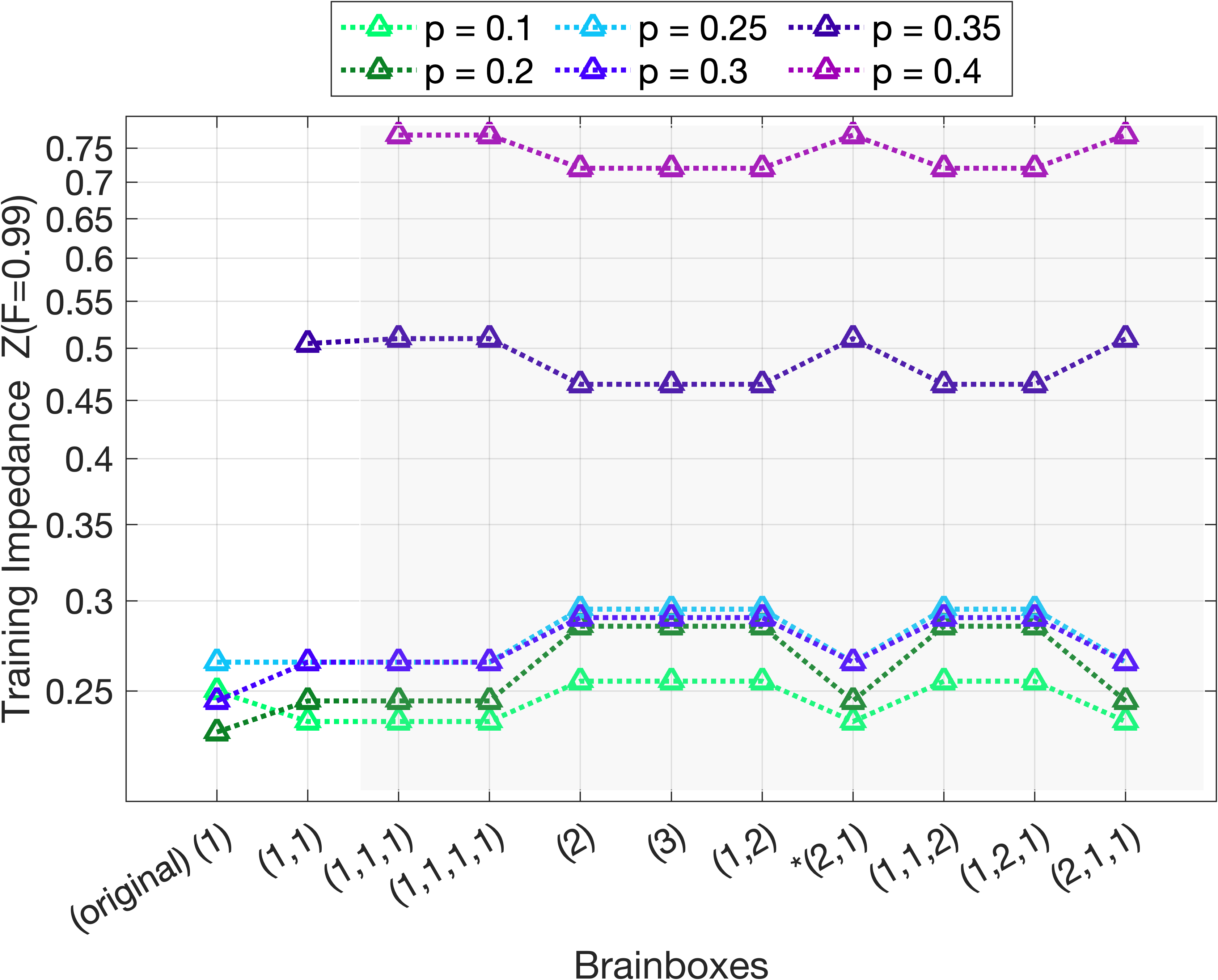}
    \caption{Training impedance for the optimization of (4,2,BB,2,4) networks. As the noise intensities grow, the optimization is more demanding. Some BBs show to be less efficient to rapidly gain high fidelity in the output state.}
    \label{fig:Trainingimpedance}
\end{figure}

Let us mark the step $n(F)$ at which the training achieves a fidelity above $F$ in the output. Consider that each BB-QAE is trained for $N_{\rm{it}}$ iterations. We define the \emph{training impedance} $Z(F) = n(F) / N_{\rm{it}}$. In Fig. \ref{fig:Trainingimpedance} we evaluate $Z(0.99)$ for several networks at different noise probabilities $p$. The result indicates that training impedance depends on the fidelity limit, training noise probability, brainbox, and input qubits, i.e.  $Z=Z(F, p, \rm{BB}, N_{\rm{in}})$. 

\begin{figure*}[t]%
    \centering%
         \subfloat[]{\includegraphics[scale = 0.18]{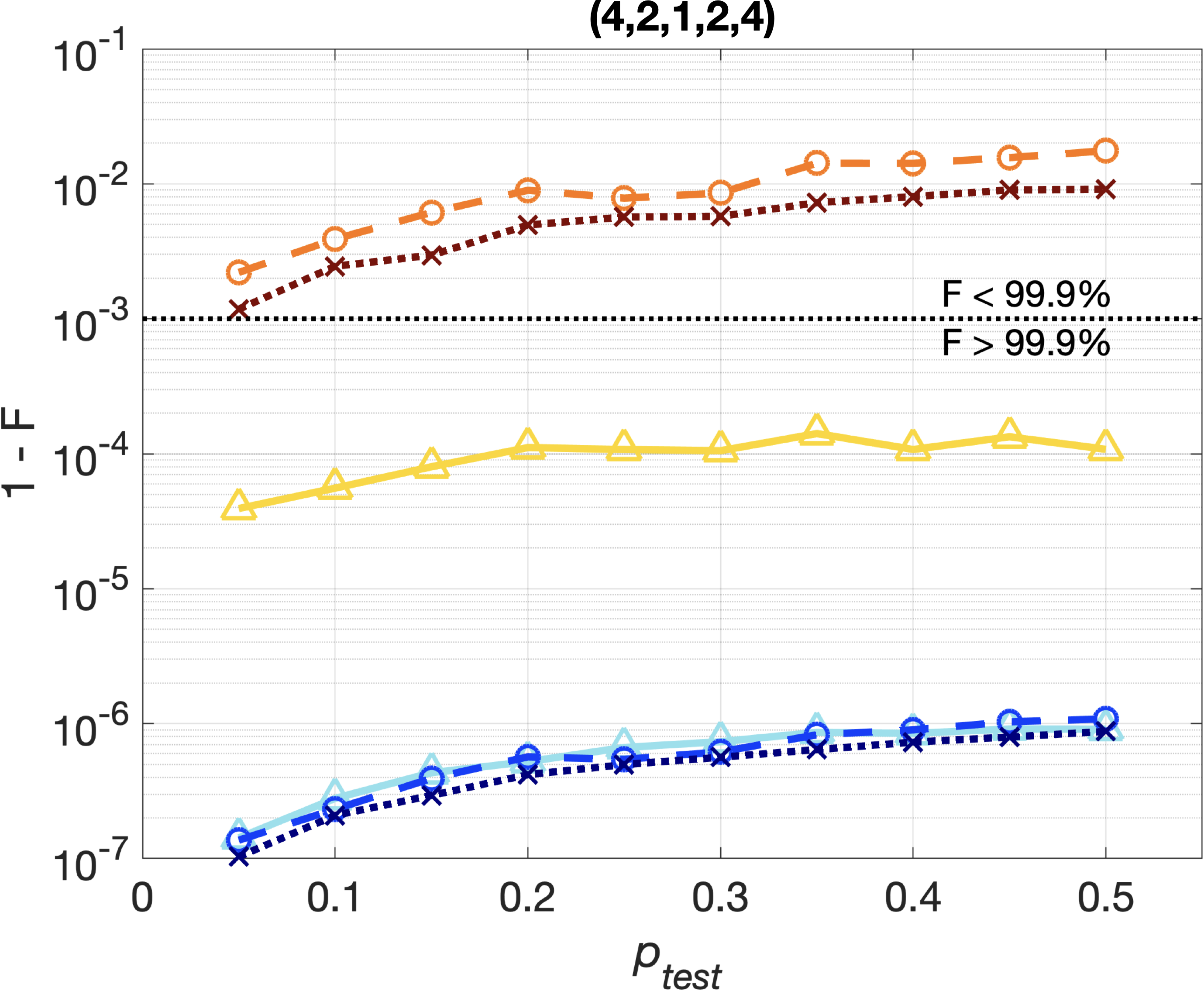}}
    \hfill
    \centering 
         \subfloat[]{\includegraphics[scale = 0.18]{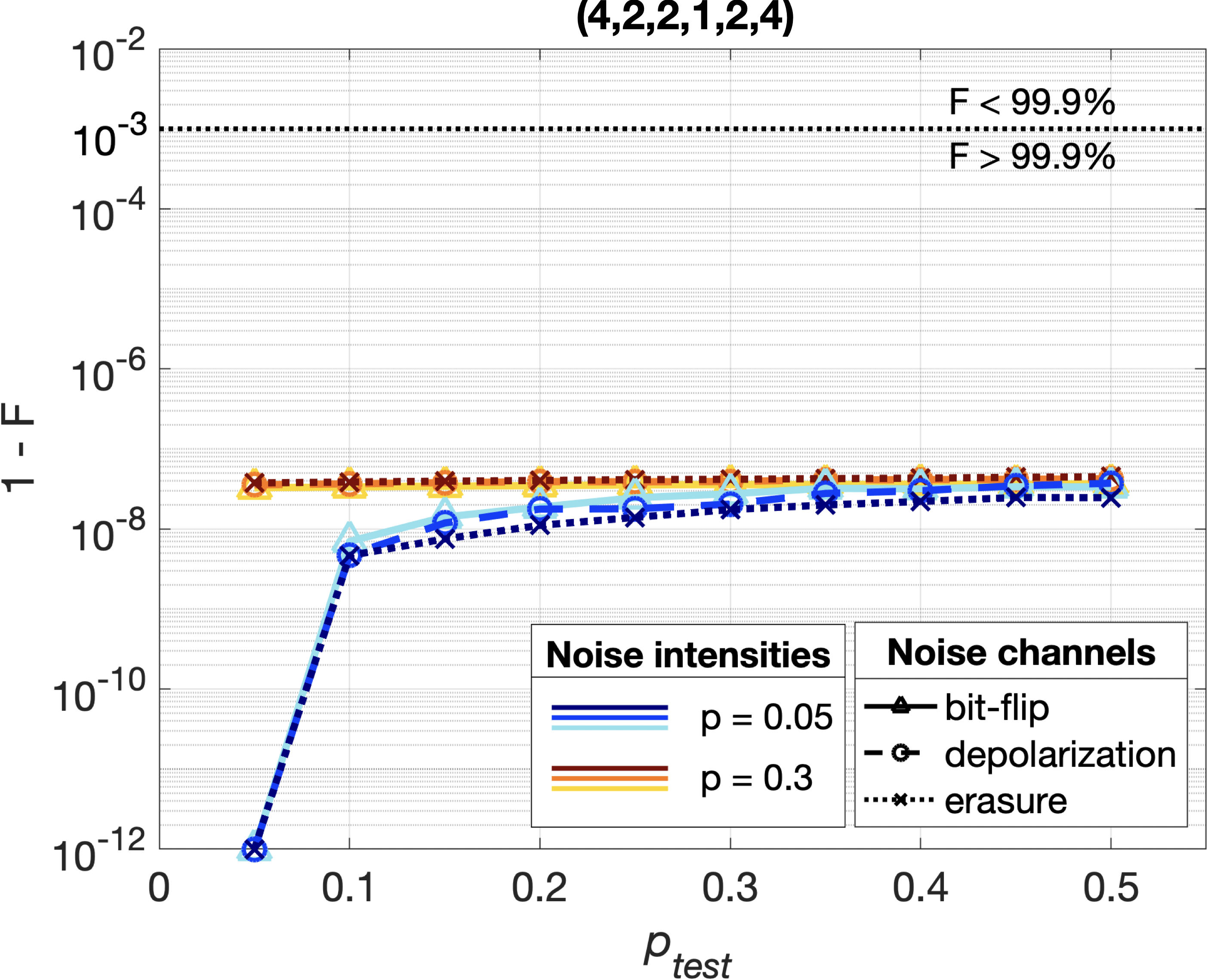}}
   	\caption{Cross-tests results for two networks: (4,2,1,2,4) and (4,2,2,1,2,4) associated to (1) and (2,1) brainbox subnetworks. Three noise channels were implemented with noise intensities $p_{test}$: the bit-flip channel (full lines), the depolarizing channel (dashed lines) and the erasure channel (dotted lines). The (1)-QAE shows more sensitivity noise in the test states: for the same training probability $p_{train}$, the reconstruction error fluctuates and larger errors occur on unfamiliar noise channels. In contrast, the map optimized by the (2,1)-QAE treats all noise channels and intensities equally. The outputs of the BB-QAE lose dependency on the noise it was trained with. In addition, the reconstruction error over the weak noise regime is lower compared to the (1)-QAE. }
	\label{fig: cross-tests}
\end{figure*}

Results for (4,2,BB,2,4) networks are summarized in Fig. \ref{fig:Trainingimpedance}.  For noise probabilities $p \leq 0.3$ the impedance factor $Z(0.99)$ in all BB networks remains relatively small nearly between 0.23 to 0.30; meaning that all networks at these noise probabilities can easily find their way to fidelity above 0.99 within the first third of the training. Some BBs such as (2), (3), (1,2), (1,1,2), (1,2,1) are slightly slower in gaining high fidelity. However the very same network under harder noise of $p\geq 0.35$ have an advantage during the training and optimization is almost 5\% faster than in other networks.

The selection of a suitable brainbox is based on the trade-off between the gain in fidelity and the loss in computational speed. At low noise intensities such as $p=0.1$, linear brainboxes (1,1),(1,1,1) and (2,1) accelerate the training compared to the single qubit box (1). Longer brainboxes also protect the network against overfitting (see section \ref{subsec:Cross-test denoising}). Between p=0.2 and 0.3, multi-qubit brainboxes cause a small computational overhead that is minimized by the linear architectures. Above the (1)-QAE's tolerance threshold, wide brainbox structures such as (2), (3), and (1,2) improve the training efficiency compared to the linear ones. Thanks to a larger amount of parameters, they efficiently capture subtle patterns in the training states, as in the over-parametrized regime \cite{Larocca2021,Rocks_Mehta_2022}.

Similar graphs for (6,2,BB,2,6) networks are shown in Appendix \ref{appendix: six-qubit results}.

\subsection{Cross-testing} \label{subsec:Cross-test denoising}

In previous sections, the testing data set was generated under the same noise channel as during the training of the quantum map. A generalization of this approach has been described in  Ref. \cite{Achache2020}, in which the QAE is trained using a noise channel with parameter $p$ and is tested with the same channel with different parameter $p'$. In this section, we evaluate the BB-QAEs with a generalized cross-test: the testing data originates either either from the same noise channel with different intensity, or from a different noise channel. 

We consider two BB-QAEs with brainboxes (1) and (2,1). Though these two brainboxes have similar impedance factors (see Fig.\ref{fig:Trainingimpedance}), they differ by their tolerance threshold (Fig.\ref{fig:Trainingimpedance}). We train them both with bit-flip noise at intensities $p_{\rm{train}}=0.05$ and 0.3.  After the training is completed, we use the final map to test noisy input GHZ states generated by one of the following three noise channels with independent noise intensities $p_{\rm{test}}$: (1) bit-flip channel defined in Eq.\ref{eq: bit-flip channel}, (2) depolarizing channel $\mathcal{E}_i^{dep}(\rho, p_{\rm{test}}) = (1-3p_{\rm{test}}/4) \rho + p_{\rm{test}}/4 (X_i \rho X_i + Y_i \rho Y_i + Z_i \rho Z_i)$, which can add a relative phase between $|00\cdots 0 \rangle$ and $|11\cdots 1\rangle$ of GHZ-states and can rotate each qubit around an arbitrary axis, and  (3) erasure channel that by probability $p_{\rm{test}}$ replaces the state of a single qubit in the GHZ state with a random state $\alpha |0\rangle + \beta |1\rangle$, otherwise it remains unchanged, \cite{Bennett1997,Grassl_Beth_Pellizzari_1997}. In the latter, since all $\alpha$'s and $\beta$'s are different for each noise realization, the map is challenged to reconstruct GHZ-states starting from any possible pure quantum state. In Fig.\ref{fig: cross-tests}, we evaluate the generalization error with the reconstruction error $R(\{\rho_x^{\rm{out}}\}, \rho_{\rm{GHZ}}) = 1 - 1/N_{\rm{test}}\sum_{x = 1}^{N_{\rm{test}}} F_x (\rho_x^{\rm{out}}, \rho_{\rm{GHZ}})$ where $N_{test} = 200$ is the number of states in the testing data set. 

For both network morphologies, training with weak noise yields almost perfect generalization to all three noise channels over a large range of probabilities. In figure \ref{fig: cross-tests}(a,b), reconstruction error is kept in the negligible range.

We repeat the same cross-testing procedure at the tolerance threshold of the (1)-QAE. In figure \ref{fig: cross-tests}(a), this network recovers from the bit-flip channel with reconstruction error close to 0.001. In contrast, states affected by the erasure and depolarizing channels cannot land on ideal GHZ state with high fidelity ( higher than 99.9\%). This is a sign of overfitting, since the discrete states in the former case are already represented in the training data set. The two remaining noise channels add states that are new to the network. In this respect, the noise tolerance measure in Fig.\ref{fig:tolerance1} is deceitful to the extend that the last optimized map works solely on the training states.

Training the (2,1)-QAE with $p_{\rm{train}}=0.3$, ie below its tolerance threshold, enables the full recovery of erroneous states irrespective of the noise channel tested, at all $p_{\rm{test}}$. This is possible due to the fact that the extended network has access to the dominating fraction of ideal GHZ states, which brings advantages in the cross-tests as well. One can think of the BB structure as a magnifying glass that makes it possible to distinguish targets from noise even when they are close to one another, by creating a better encoding of inputs in its last layer.

\subsection{R\'enyi entropy flow}

A key property to measure in engineered quantum systems is entanglement: in contrast to their classical counterparts, quantum algorithms can generate large amounts of entanglement between parts of the system \cite{PhysRevLett.91.147902,RevModPhys.82.277}. Entanglement during the learning phase in a QAE changes internally across layers. It allows delocalizes information in the network and steers the training towards the optimal condition of having a separable output. In order to observe its contribution to the training, some measures of entanglement have been tested, such as entanglement witnesses \cite{Brown2022} and von Neumann entropy \cite{Ballarin2022}. Similar to any many-body quantum system, measuring the entropy of different partitions provides a way to probe its entanglement structure. 

Here, we evaluate the second-order R\'enyi entropy since it can capture long-range entanglement  \cite{Ansari_2019,Ansari2015_exact_correspondence,Ansari2015_entropy_heat_engines} as well as dissipation mechanisms \cite{Uzdin2021,Nazarov-parallel}.  R\'enyi entropy can serve as a measure for probing and characterizing brainbox bottlenecks.  A slow entropy growth in a layer or in a part of the network can be used to identify localization in a subset of the network \cite{RevModPhys.91.021001}.

\begin{figure*}[t]%
    \centering%
         \subfloat[][Network (4,2,1,2,4)]{\includegraphics[scale = 0.3]{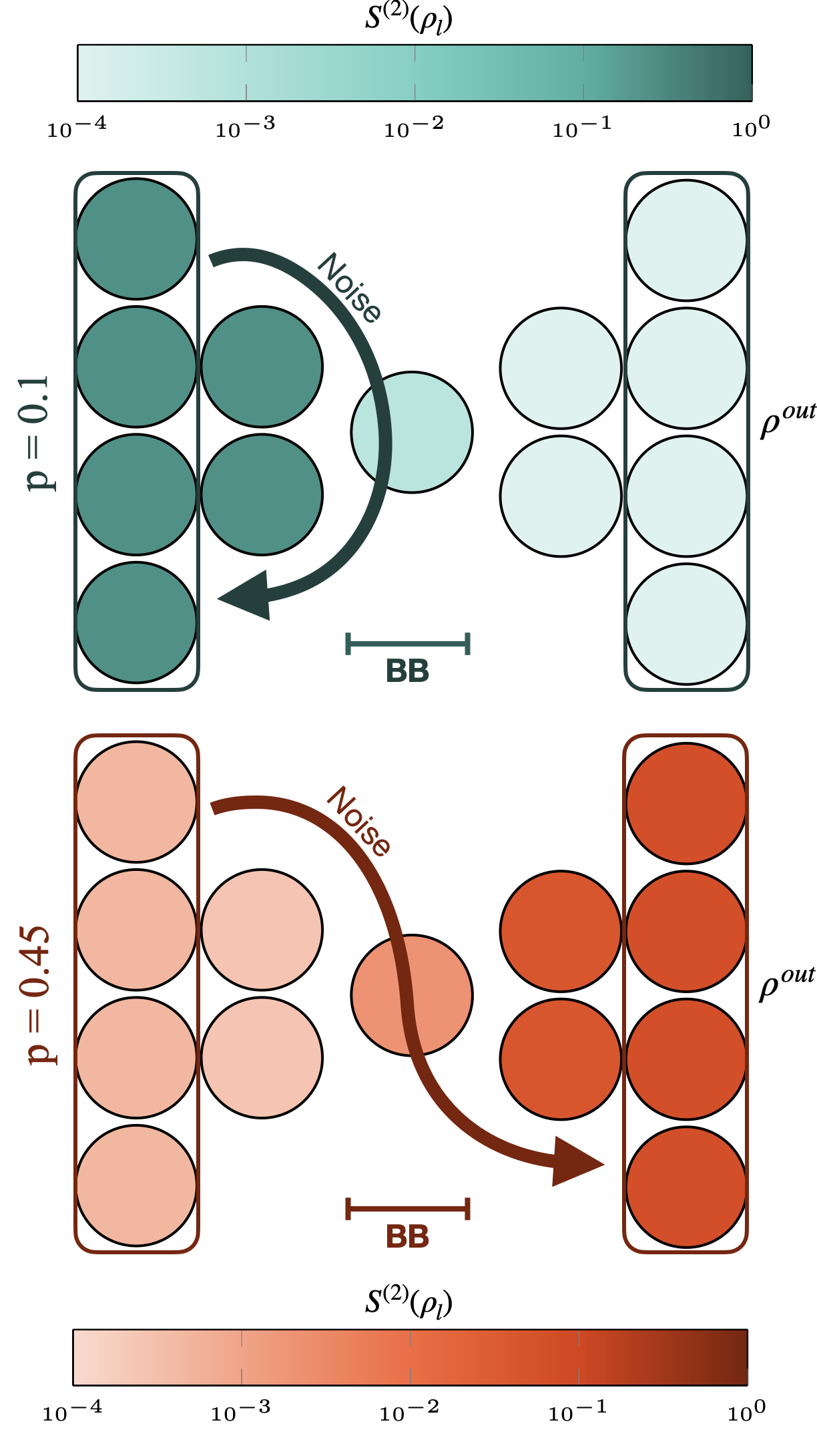}}\ \ \ \ \ \ \ 
    \centering 
         \subfloat[][Network (4,2,2,2,4)]{\includegraphics[scale = 0.3]{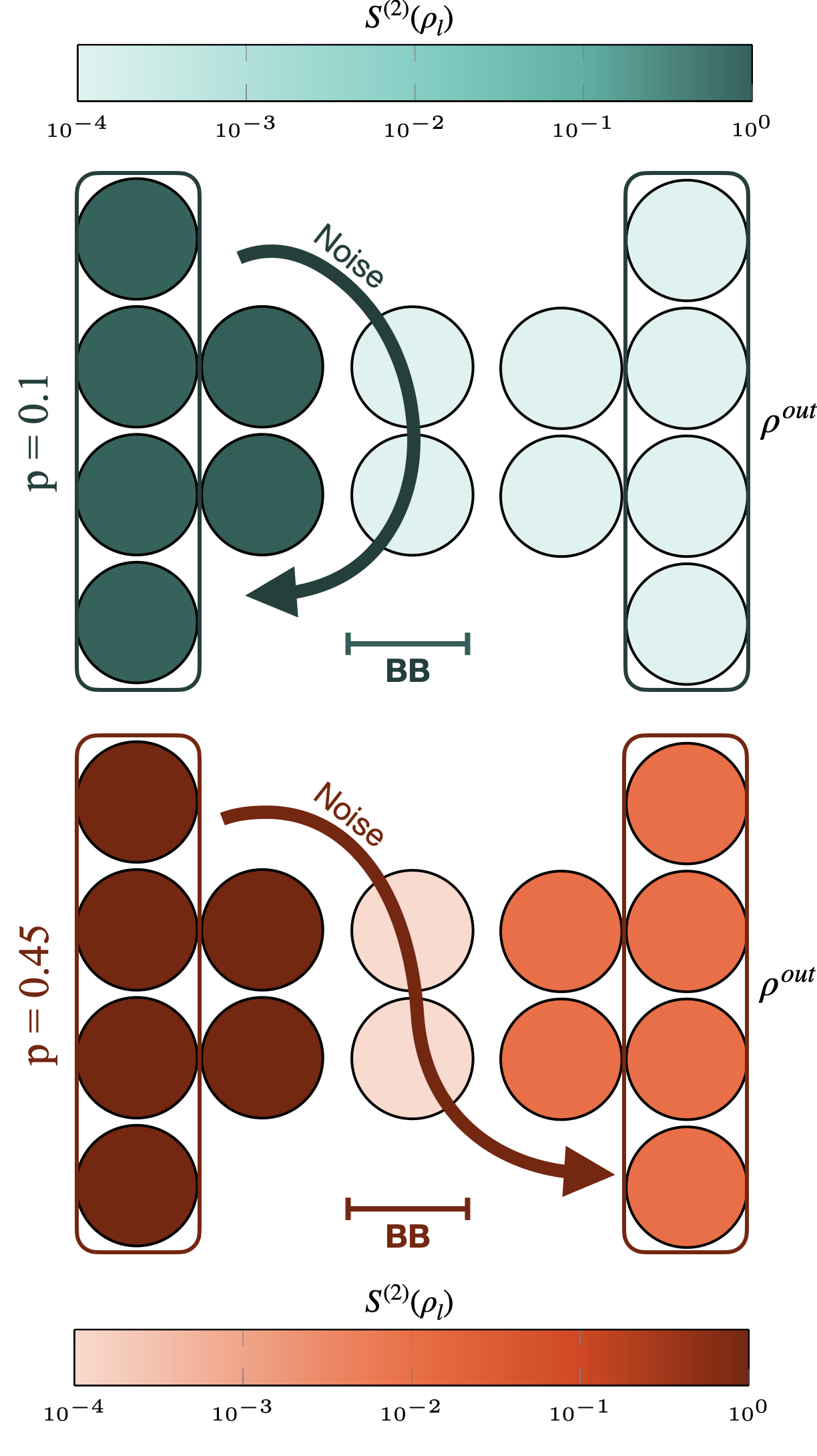}}
   	\caption{Layerwise R\'enyi entropy evaluated after applying the last step of the denoising map. Darker colors indicates larger entropy of noisy mixed states. We study a single-qubit brainbox in (a) and a double-qubit layer brainbox in (b). Entropy at low noise strength of $p=0.1$ decreases toward the output layer so that one can expect noise is localized in the encoder and is blocked away from the brainbox. However in the case of $p=0.45$ which is much stronger than  the networks' tolerance threshold, input noise from the input layer leaks out at the bottleneck and noise accumulates in the decoder and output layer. }
	\label{fig: entropy 42124 42224}
\end{figure*}

For a bipartite system $\mathcal{S}$ with subsystems A and B and total density matrix $\rho$, second order R\'enyi entropy is $S^{(2)}(\rho)=- \log{(\Tr\{\rho^2\})}$. When equal to zero, it indicates that $\mathcal{S}$ is pure and independent from any environment. Typically, entropy of the whole BB-QAE is zero at all iterations because the system is isolated from the environment and therefore in a pure state. Moreover, second order R\'enyi entropy can be evaluated for any subsystem in $\mathcal{S}$, eg. A,  based on the associated partial density matrix  $\rho_A = \Tr_B (\rho)$: $S_A^{(2)}(\rho_A) = - \log{(\Tr_A\{(\rho_A)^2\})}$. Consequently, at each training step, in a BB-QAE with $L$ layers, the entropy of layer $l$  reflects the presence of entanglement between the layer $l$ and the remaining $L-1$ layers in the network. The second order R\'enyi entropy in layer $l$ is defined as
\begin{equation}
S_l^{(2)} = - \log{(\Tr_l\{(\rho_l)^2\})},
\label{eq: Renyi entropy}
\end{equation}
with the partial density matrix of layer $l$ being $\rho_l = \Tr_{k \neq l}\{\rho\}$ for $k=1,\cdots L$ and $\rho$ is the state of the whole BB-QAE. 

 In particular, at each iteration, the entropy of layer $l$ can be evaluated using  Eq.(\ref{eq: Renyi entropy}) after applying the respective unitary $\mathcal{U}^l$. During the training, we compare the evolution of layer-wise entropy in a (1)-QAE for both weak ($p=0.1$) and strong ($p=0.4$) noise in the input GHZ states (see Fig.\ref{fig: entropy evolutions} in Appendix \ref{App: entropy evolutions}). During the learning phase, entropy is redistributed within the network. In the first steps, it undergoes steep growth, especially in the last layer. In the subsequent iterations, entanglement vanishes exponentially in the decoder's layers, while it is only slightly suppressed in the encoder, resulting in entropy inversion.

Entropy after optimization is compared for (1)- and (2)-QAEs below and above the tolerance threshold, at $p=0.1$ and $p=0.45$ respectively (Fig.\ref{fig: entropy 42124 42224}). In a BB-QAE with bit-flipped GHZ state on the initial layer, successful denoising not only raises fidelity of the output states, but also improves its separability. Therefore, training inverts entropy in the network and shifts noise from the decoder to the encoder. The bottleneck seals it away from the output layer.

In contrast, failure to denoise the inputs can take two forms. In Fig.\ref{fig: entropy 42124 42224}(a), instead of concentrating noise in the encoder, the training yields high entanglement in the last two layers, while the encoder remains almost independent. As in Fig.\ref{fig: entropy 42124 42224}(b), the inversion of entropy can be favorized by using larger BB structures. In this case, the training improves noise concentration, but the bottleneck seal seems too porous to lock noise out of the decoder, resulting in poor denoising.

\section{Conclusion}

We have presented an in-depth study of various brainbox structures for the bottleneck in a quantum autoencoder used to denoise entangled quantum states. Training a QAE single-qubit bottleneck has been studied in Ref. \cite{Bondarenko2020}. This bottleneck can come with only limited tolerance against bit-flip, depolarizing, and random unitary noise channels. Scaling the inputs size from 4 to 6 qubits makes the training more greedy in data, and deteriorates the denoising performance rapidly. 

We identified two mechanisms behind the limitation of noise tolerance. (1) The finite size of the training data set causes statistical deviations from the ideal noisy state distribution expected from the bit-flip channel. It imposes an upper bound on the maximum tolerance the BB-QAE can achieve. This upper bound depends on each training data set. (2) The study of R\'enyi entropy shows that the single-qubit bottleneck is unable to seal noise away from the output state, and therefore to carry out its denoising task.

We compared the simple QAE with multi-qubit brainbox bottlenecks, most of which brought significant elevation of tolerance. When qubits are added to the input and output layers, the relative improvements are maintained. If a brainbox bottleneck can endure stronger noise  compared to another brainbox, adding more qubits to input state maintains the superiority of the former one. 

Some bottlenecks show similar tolerance threshold against noise. This raises an important question: What other features can make a brainbox more suitable than the other ones?  To address this question, we compare training impedance between brainboxes. For this purpose, we evaluate the training impedance $Z(0.99)$, which indicates what minimum percentage of the training process is required to achieve a fidelity above 99\% in the output. The result has been summarized in Fig. (\ref{fig:Trainingimpedance}) and show that the training impedance depends not only on the bottleneck, but also on the training noise probability $p$. Below bit-flip probabilities $p=0.3$, linear brainboxes such as (1,1) are favorable to a more efficient training. In contrast, between $p=0.3$ and $p=0.4$, non-linear brainboxes such as (2) or (2,1) are most economical to train.

We evaluate the R\'enyi entropy of network layers at each optimization step to show how nonlocal entanglement between layers evolves and impacts the outputs fidelity. Results show that in networks below their tolerance threshold, entropy becomes localized in the encoder of the BB-QAE, so that much less noise passes through the bottleneck to the decoder. This usually leads to outputs states that have high fidelity with the target and that are separable from the network. Some examples were given in Fig.\ref{fig: entropy 42124 42224}: in successful training, noise is blocked off from the bottleneck, while in unsuccessful training noise penetrates through the bottleneck.  The absence of separability of the output indicates the presence of layer-to-layer stray coupling between hidden and output layers, which  eventually does not allow its fidelity to rise higher.

In connection to NISQ devices, QAEs are resilient to input layer noise and therefore they provide the potential to generate ideal entanglement on noisy gates and qubits. A QAE with complex bottleneck and more qubits and parameters in general seem advantageous for denoising, because such a complex structure provides the possibility to separate encoder and decoder. However detailed analysis shows that less resourceful brainboxes can be found with the same performance as a complex one. Testing the network with the depolarizing and erasure channel proves that some bottlenecks can keep their superiority over the whole trainable range. We expect that these differences will remain when selecting different quantum target states.

One of the main obstacles against implementing QAEs in scaled up input states is the required high connectivity in the network that is inaccessible on the current processors. An alternative is to train a map with missing connections \cite{Bondarenko2020}.

\section*{Acknowledgement}
The authors thank Maria Schuld and Pia D\"oring for fruitful discussions. MA acknowledges that a part of this manuscript was motivated during the support from Intelligence Advanced Research Projects Activity (IARPA) under contract W911NF-16-0114.

\bibliography{Paper_QAE.bib}

\newpage
\onecolumngrid

\appendix

\section{Denoising a 6-qubit input GHZ states on (6,2,BB,2,6) networks}
\label{appendix: six-qubit results}

In this section we present results related to denoising 6-qubit input GHZ states.

\subsection{Training procedure}

\begin{figure}[h!]
    \centering
   {\includegraphics[scale = 0.47]{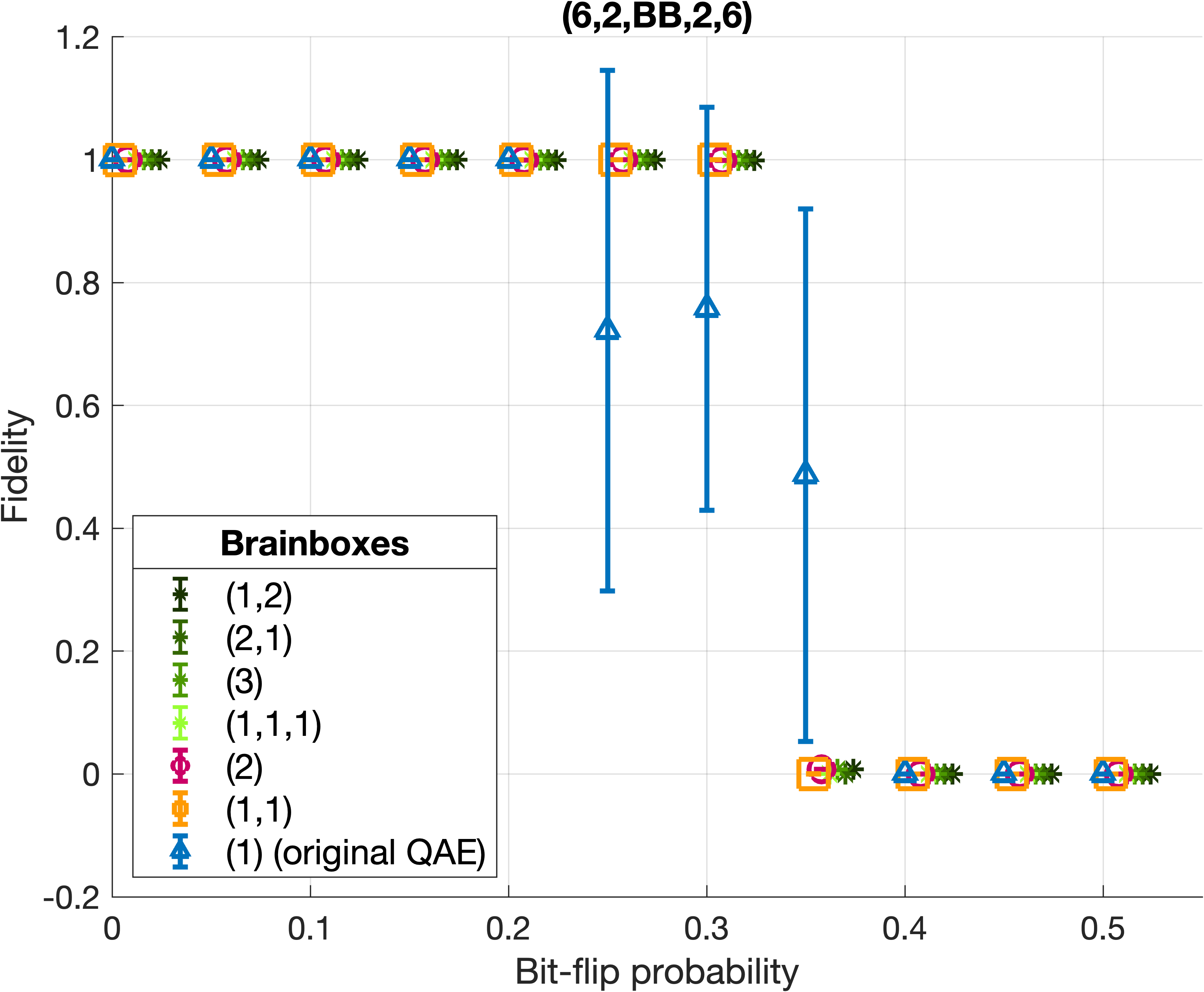}}    
     {\includegraphics[scale = 0.53]{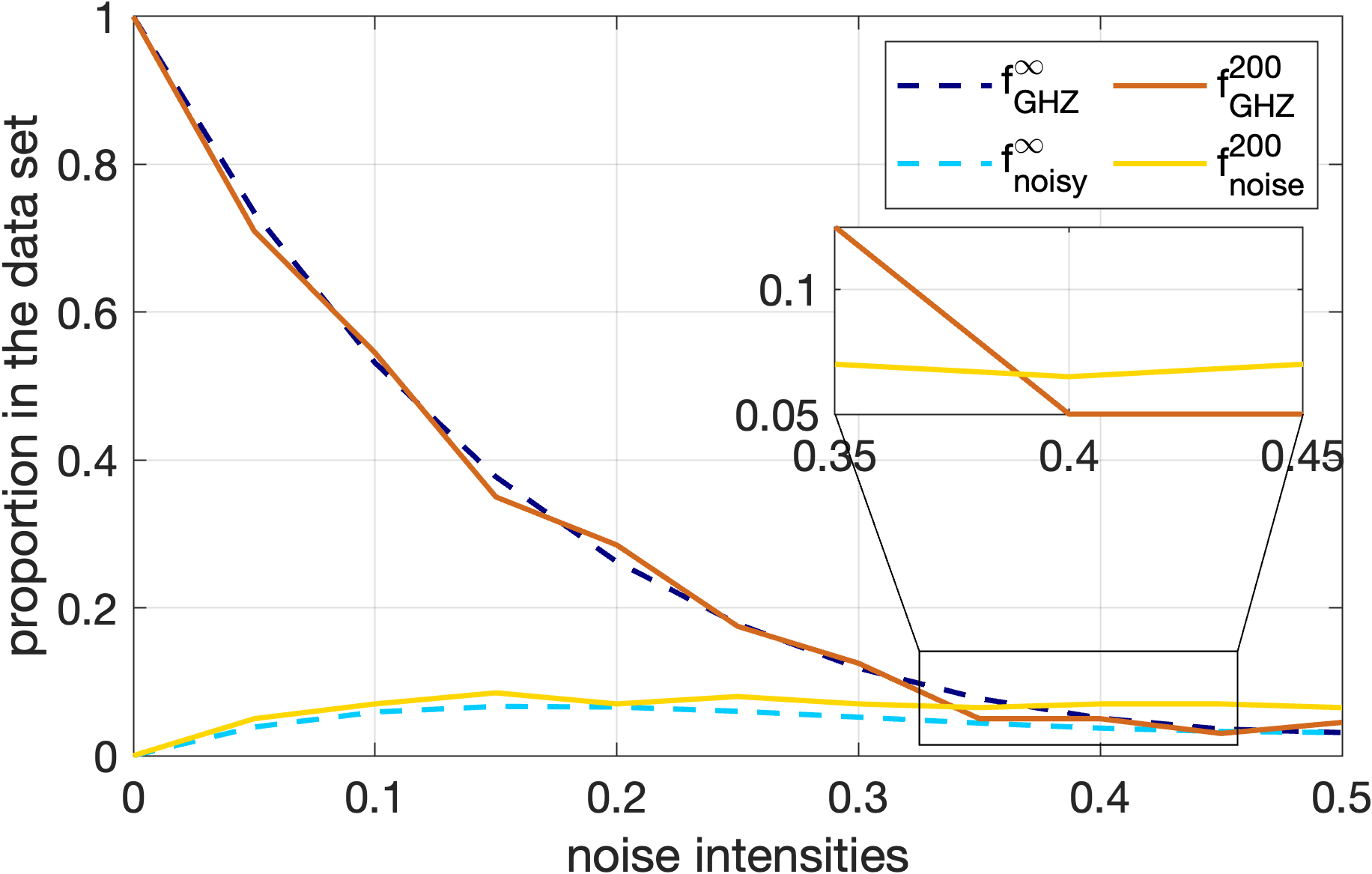}}
    \caption{(a) Testing fidelty: Average output state fidelity over a range of noisy input test states with noise probability $p$. The error bars indicates the absolute value of standard deviation in the data about average fidelity. (b) Training data set: The distribution of 6-qubit GHZ and non-GHZ states in infinitely many samples (dashed line) versus finite 200 samples (solid lines). In the infinite sample case the distribution of GHZ for all noise probabilities $p$ dominates, while artefacts in the finite data set prevents the dominance at strongly noisy channels.}
    \label{fig:app_fidelity}
\end{figure}

The input layer is initialized with 6-qubit inputs, while the rest of the network is initialized in the ground state. The network is trained with a set of 200 noisy GHZ states and bit-flip probabilities $p$. After the training is over, the optimized quantum map is used to test the performance on some new noisy GHZ states the network was not trained with. The result is repeated for 200 test states at different $p$ values between 0 and 0.5. For every choice of brainbox, we evaluate the output state fidelity. Results are plotted in Fig.\ref{fig:app_fidelity}.

As discussed in the main text, in the limit of infinite size data set, the distribution of GHZ and non-GHZ states is such that the amount of GHZ states is always larger than that of non-GHZ except at $p = 0.5$.
This distribution in Fig.\ref{fig:app_fidelity}(b) is shown in dashed line. Reducing the size of our training data to a finite value makes zigzag deviations about the ideal distribution. Around $p=0.35$, the finiteness disorder reverts the superiority of GHZ state which deceives the network into recovering an undesired target state. Solid lines in Fig.\ref{fig:app_fidelity}(b) shows the training data we used to denoise a 6-qubit QAE.

\subsection{Training impedance}

\begin{figure}[h!]
    \centering
    {\includegraphics[scale = 0.6]{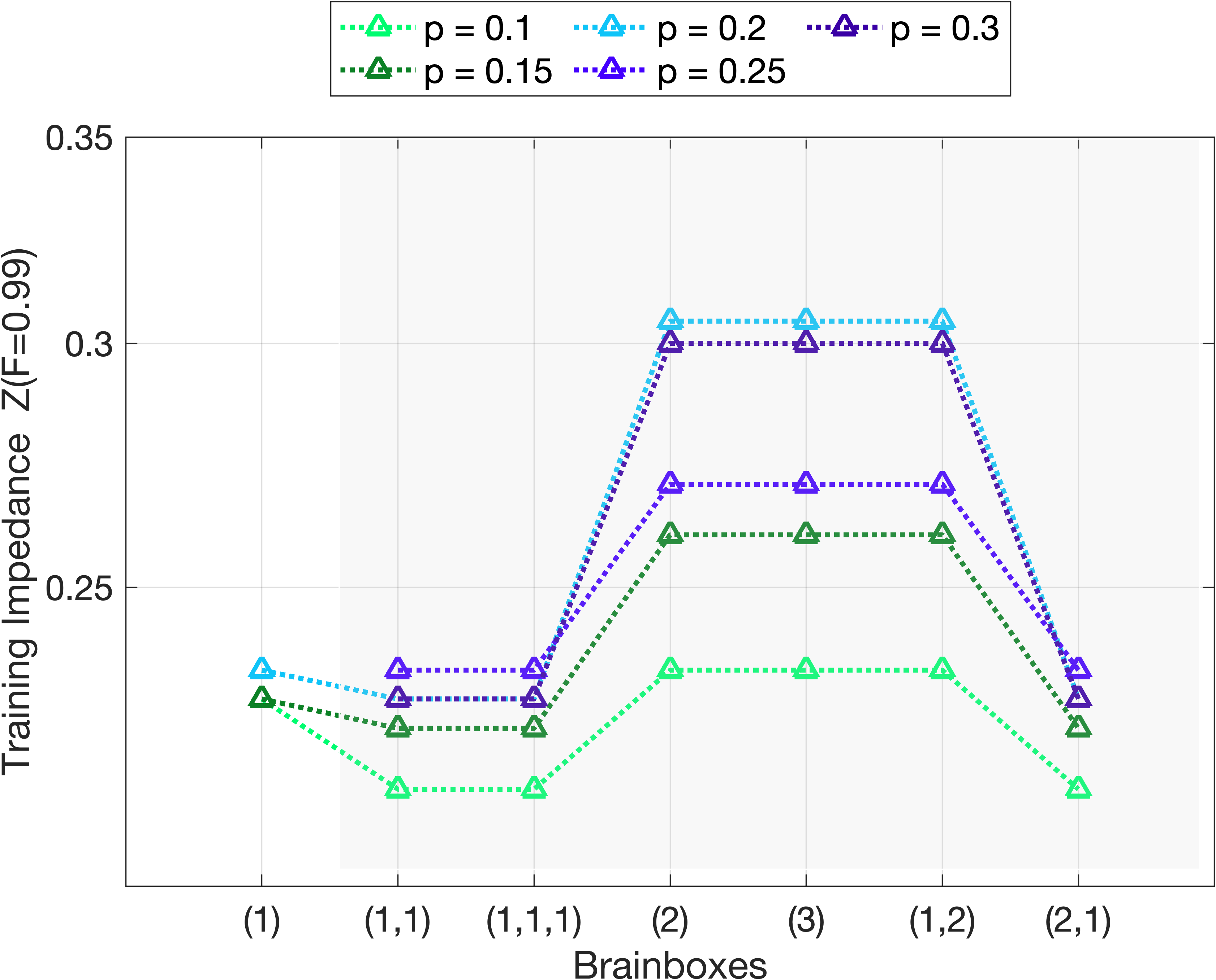}}    \caption{Training impedance for the optimization of
(6,2,BB,2,6) networks. In contrast to the 4-qubit inputs, linear BBs perform best at all noise intensities, except for the (2,1)-configuration.}
    \label{fig:app_impedance}
\end{figure}

Results for (6,2,BB,2,6) networks listed in Fig.\ref{fig:app_fidelity}(a) show the tolerance threshold for denoising outputs. However, some complex BBs such as (1,2), (3) reach the same tolerance than simpler BBs, such as (1,1), (2). In order to understand which BB is more efficient, we evaluate training impedance in the networks. Results can be found in Fig. (\ref{fig:app_impedance}). For all $p$ values within the range indicated in the plots, the network (1,1) and (1,1,1) have less resistivity against training compared to the network (2) and (3), and even (1,2).

\subsection{Cross testing}

In the section the result of cross testing of the (6,2,1,2,6) network with brainbox (1) is discussed. As mentioned in the previous appendix sections, the (1)-QAE network can tolerate noise in the domain of $p\leq 0.2$. In this range of noise strengths, the network carries an intermediate impedance to training, which makes it efficient for training. Training the network with bit-flip noise channel of $p_{\rm{train}} = 0.05$ trains the network based on a dominant subset of GHZ states in the training data set.

\begin{figure}[h!]
    \centering
    {\includegraphics[scale = 0.5]{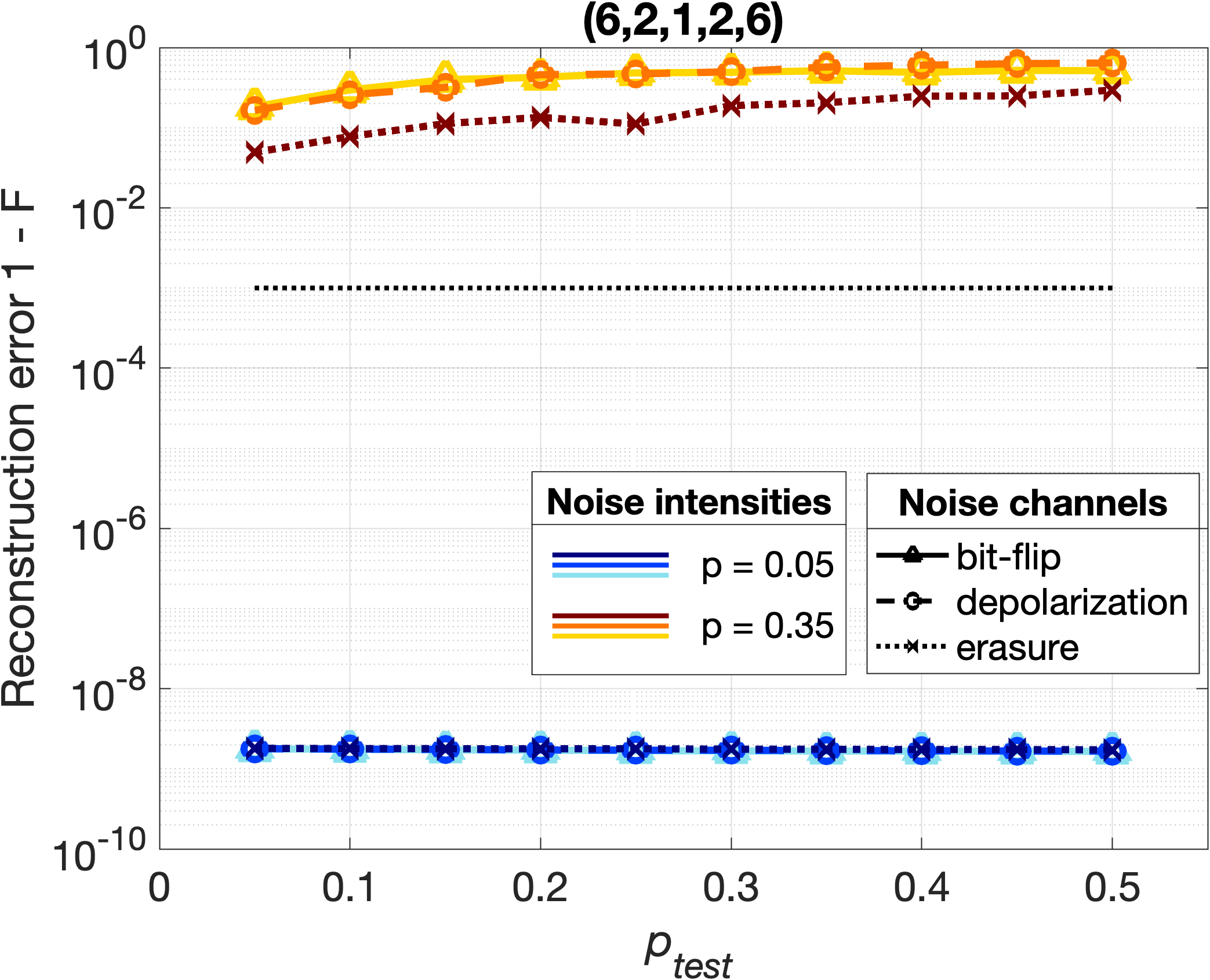}}    \caption{Cross-tests results for two networks: (6,2,1,2,6) associated to the brainbox subnetwork (1). Three noise channels were implemented with noise intensities $p_{\rm{test}}$: the bit-flip channel (full lines), the depolarizing channel (dashed lines) and the erasure channel (dotted lines).}
    \label{fig:app_cross}
\end{figure}

This training makes the network resilient to (1) bit-flip, (2) depolarizing, (3) erasure channels in a large domain of noise strength $p_{\rm{test}}<0.5$. However training the network with input noise probability beyond the network tolerance makes the network confused about the identity of the dominant subset in the training data set. This suppresses the fidelity of output state to $<90\%$.

\section{Entropy evolution} \label{App: entropy evolutions}

\begin{figure}[h!]
    \centering
  (a)  \includegraphics[scale=0.85]{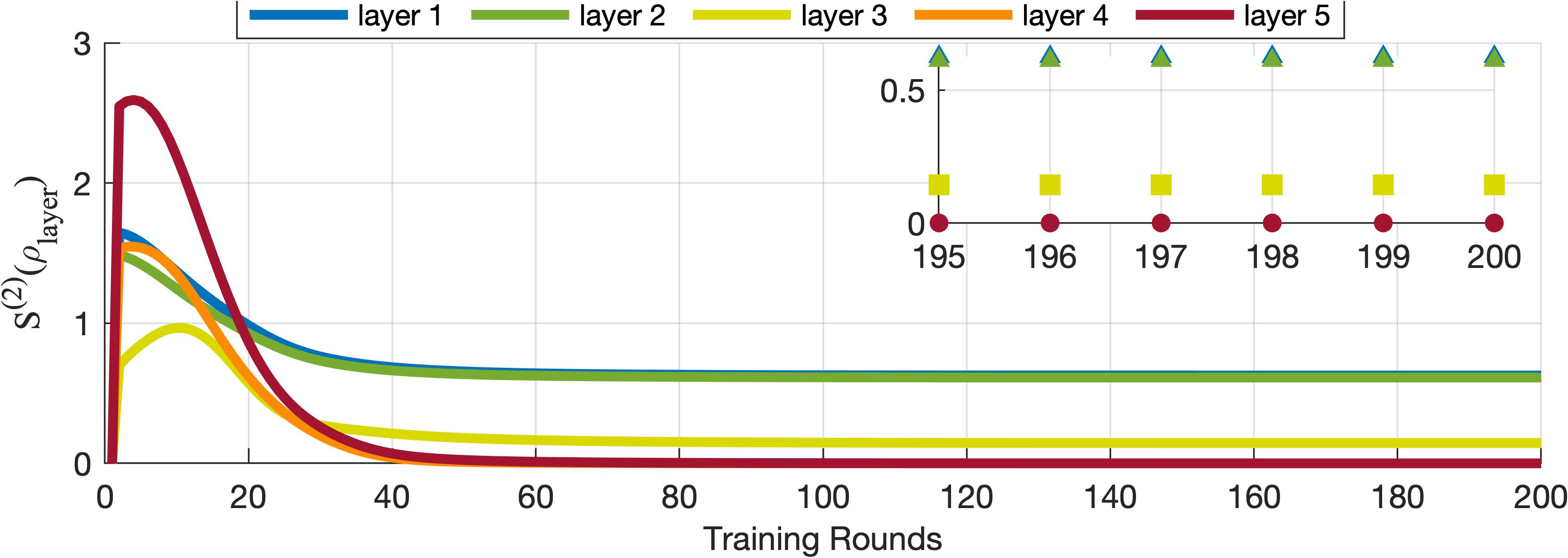}\\
 (b)    \includegraphics[scale=0.85]{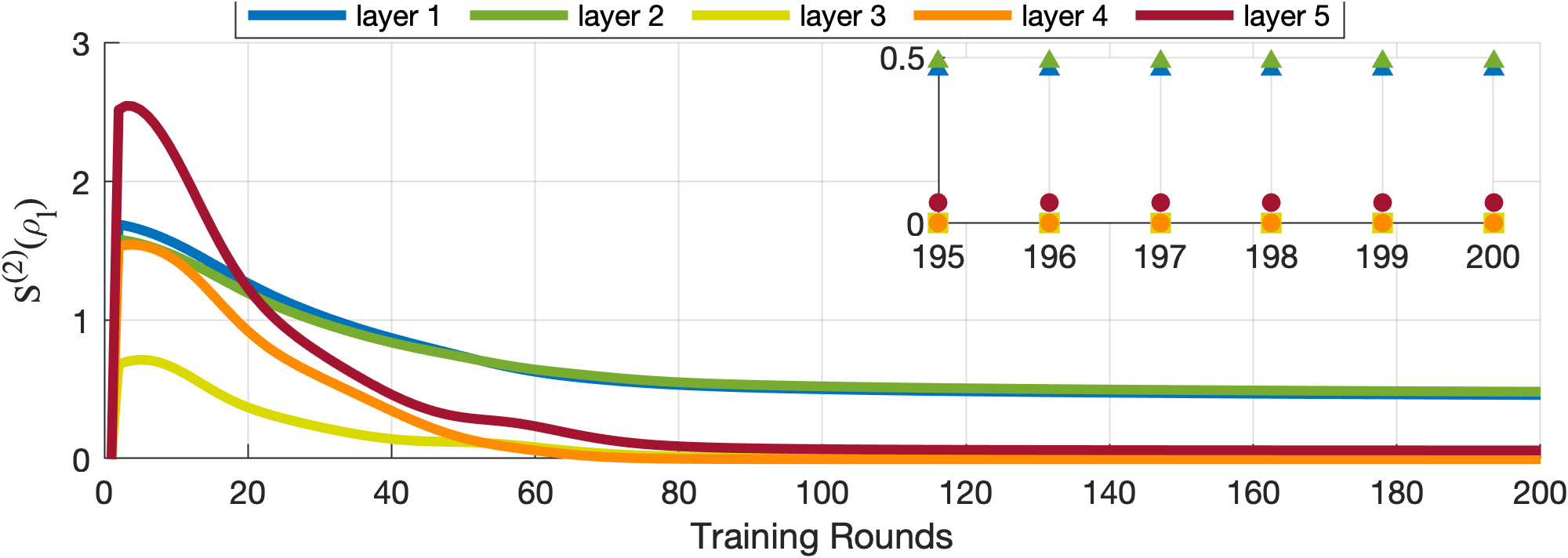}
    \caption{R\'enyi entropy flow in individual layers of (4,2,1,2,4) network during its evolution during the training (x-axis). The bit-flip probability $p$ of the noisy input GHZ states is  $0.1$ in (a) and 0.4 in (b).}
    \label{fig: entropy evolutions}
\end{figure}

In this appendix we list some result on the time evolution of entropy during  training steps. We consider the network (4,2,1,2,4) with (1)- bottleneck.  On this network we start the input layer (layer 1) with a set of noisy GHZ states with noise strength $p$. All other qubits in other layers are in the ground state.  We initialize a quantum map at random and optimize it iteratively to create ideal GHZ states on the output layer (layer 5). At each step we evaluate total network density matrix and by tracing out the irrelevant layers, we evaluate the second order R\'enyi entropy for each layer. The result for $p=0.1$ can be seen in Fig. (\ref{fig: entropy evolutions} a) and for $p=0.45$ in Fig. (\ref{fig: entropy evolutions} b).

As expected, all layers start from zero entropy and quickly raise their entropy as they capture mixed state from the noisy input. Even as one can see the output layer (layer 5) shows a large entropy after a few steps of optimization. 

Continuing optimization lowers the entropy of the decoder (layers 4 and 5) much faster than in the encoder (layers 1 and 2). The end of the denoising processes have been magnified in the two insets in (a) and (b). One can see that in the weak noise regime of (a) with $p=0.1$ layer 5 carries zero entropy, which makes it a separable state from other layers. In the strong noise regime (b)  with $p=0.45$, entropy of the output layer is finite and larger than that of the bottleneck (layer 3), which makes the state entangled to other layers and therefore being affected by input noise. This prevents this network to land on stable high fidelity GHZ state due to stray couplings.

\end{document}